\documentclass[nohyper,notoc]{JHEP3}
\usepackage{amsmath,amssymb,amsfonts,epsfig,dsfont}
\usepackage{latexsym}
\usepackage{graphicx}
\usepackage{subfig}

\def\bse{\begin{subequations}}
\def\ese{\end{subequations}}

\def\a{\alpha}

\def\m{\mu}

\def\Im{{\rm Im}}

\newcommand{\Tr}{{\rm Tr\,}}
\newcommand{\half}{{\textstyle{\frac12}}}

\newcommand{\beq}{\begin{equation}}
\newcommand{\be}{\begin{equation}}
\newcommand{\ee}{\end{equation}}
\newcommand{\bea}{\begin{eqnarray}}
\newcommand{\eea}{\end{eqnarray}}
\renewcommand{\Re}{\mathrm{Re}\,}
\renewcommand{\Im}{\mathrm{Im}\,}
\newcommand{\pa}{\partial}
\newcommand{\pab}{\bar{\partial}}

\newcommand{\nn}{\nonumber}
\newcommand{\zb}{\bar{z}}
\newcommand{\ub}{\bar{u}}
\newcommand{\betab}{\bar{\beta}}
\newcommand{\vv}{{\mathbf v}}
\newcommand{\Rb}{\bar{R}}
\newcommand{\As}[1]{\,_{#1}\!A}
\newcommand{\Abs}[1]{\,_{#1}\!\bar{A}}
\newcommand{\Ab}{\bar{A}}

\title{Minimal surfaces in AdS space and Integrable systems}

\author{Benjamin A. Burrington\\
Department of Physics, University of Toronto,\\ 60 St. George st, Toronto, ON M5S 1A7, Canada\\
{\tt E-mail: benburri@physics.utoronto.ca}
}

\author{Peng Gao\\
Department of Physics, University of Toronto,\\ 60 St. George st, Toronto, ON M5S 1A7, Canada\\

Perimeter Institute for Theoretical Physics,\\
Waterloo, Ontario, N2L 2Y5, Canada\\
{\tt E-mail: pgao@perimeterinstitute.ca}
}

\abstract{We consider the Pohlmeyer reduction for spacelike minimal area worldsheets in AdS$_5$.  The Lax pair for the reduced theory is found, and written entirely in terms of the ${\rm A}_3={\rm D}_3$ root system, generalizing the ${\rm B}_2$ affine Toda system which appears for the AdS$_4$ string.  For the ${\rm B}_2$ affine Toda system, we show that the area of the worlsheet is obtainable from the moduli space K\"ahler potential of a related Hitchin system. We also explore the Saveliev-Leznov construction for solutions of the ${\rm B}_2$ affine Toda system, and recover the rotationally symmetric solution associated to Painleve transcendent. }

\preprint{PI-STRINGS-167\\ arXiv:0911.4551}


\begin{document}

\section{Introduction}

The gauge/gravity duality, and AdS/CFT in particular \cite{Maldacena:1997re}, offers new tools for understanding the strong coupling dynamics of certain field theories.  The most well understood examples are strictly AdS/CFT, having a product geometry with one AdS factor \cite{Maldacena:1997re,Kachru:1998ys}.

In AdS/CFT certain quantities in the dual field theory can be calculated using the semiclassical limit of the string worldsheet theory.  In the canonical example of AdS$_5 \times $ S$^5$, semiclassical worldsheets solutions are used to describe high spin operators \cite{Gubser:2002tv}, expectation values of Wilson loop operators \cite{Maldacena:1998im, Rey:1998ik}, and more recently, universal infrared characteristics of the MHV gluon scattering amplitudes \cite{Alday:2007hr} in the dual field theory.  The study of spacelike classical worldsheets in pure AdS has been revitalized since the work of \cite{Alday:2007hr, Alday:2007he}.  In \cite{Alday:2007hr, Alday:2007he} they showed that the leading contributions to the scattering amplitude is dominated by a worldsheet ending on a null polygon at the boundary of a  T-dual AdS space, finding agreement with the BDS ansatz \cite{Bern:2005iz}.  At the level of amplitudes, this T-duality  between AdS spaces was previously noticed in field theory as a new conformal symmetry realized in momentum space \cite{Anastasiou:2003kj, Drummond:2006rz}. For a recent review on both sides of the story see \cite{Alday:2008yw}.

In \cite{Alday:2007hr} the regulated area of the classical worldsheet yielded the value of the scattering amplitude, therefore it is desirable to find an efficient way to evaluate the worldsheet area for given boundary conditions.  The Pohlmeyer reductions \cite{Pohlmeyer:1975nb} of classical worldsheets in AdS offer a natural framework for this: these reduced systems carry less information and are easier than solving for the exact embedding profile of the worldsheet in AdS. However, they still contain the needed information about the area of the classical worldsheet.

The key observation is that such reduced worldsheet theories give rise to classically integrable systems \cite{Pohlmeyer:1975nb}, and so a variety of analytical techniques can be brought to bear.  The integrability of these theories may not be surprising, considering that the classical worldsheet theory by itself is an integrable system (for AdS$_n$), in the sense that it admits an associated linear problem, a Lax pair.  Because the Pohlmeyer reduced system is derived from this original integrable AdS$_n$ system, one may hope that the integrability descends as well.  One could argue that one in fact {\it expects} the integrability, given that the dressing technique acts transitively on the space of solutions of the AdS$_n$ theory, and so one expects these maps from solutions to solutions to descend to the reduced theories as well.

The Pohlmeyer reduction has already been used for studying classical and quantum aspects of string worldsheets quite extensively (see, for example, the work of \cite{Barbashov}).  It was applied quite a while ago \cite{DeVega:1992xc} to find classical worldsheets in de Sitter space.  More recently, in \cite{Grigoriev:2007bu}  the Pohlmeyer reduction of the full fledged AdS$_5 \times $S$^5$ superstring was explored, in \cite{Roiban:2009vh} the UV finiteness of the reduced theory was verified to 2-loop order, and in \cite{Hoare:2009rq}  fluctuations about classical solutions of the reduced model are studied.  In related developments, \cite{Miramontes:2008wt} obtained Pohlmeyer reductions for general coset space sigma models, and in \cite{Hollowood:2009tw} they investigate the soliton transformations of such models.  Detailed studies of $SO(3)/SO(1)$ coset models were carried out in \cite{Bakas:1995bm}.  In aspects more related to the scattering amplitudes, in \cite{Jevicki:2009bv} radial  solutions to the Pohlmeyer reduced equations are obtained using approximate series expansions, in \cite{Dorn:2009kq} they study various equations governing both timelike and spacelike minimal surfaces, and \cite{Dorn:2009gq} considers the AdS$_4$ case of spacelike minimal surfaces bound by null polygons in some detail, and \cite{Ryang:2009ay} looked at more details of solutions to the linear problem in AdS$_3$.  Finally, we note that the original version of our work was released on the same day as the related work of \cite{Alday:2009dv}.

The most important guideline for us came from \cite{Alday:2009yn}, where they restrict to considering worldsheets embed in AdS$_3$ subspaces of AdS$_5$ ( detailed AdS$_3 \times$S$^1$ embeddings may be found in \cite{Sakai:2009ut}). Due to the higher amount of symmetry in the kinematically constrained system,  they were able to show that the worldsheet area can be computed indirectly by finding the K\"ahler potential for the (hyperK\"ahler) metric on the moduli space of a Hitchin system. Moreover, this Hitchin system arises from precisely the linearization of the integrable set of equations obtained by Pohlmeyer reduction.  Using results due to \cite{Ooguri:1996me, Seiberg:1996ns}, the new eight gluon amplitude was obtained and by relating to the recent systematic study \cite{Gaiotto:2009hg} of wall-crossing behavior of BPS spectrum in Seiberg-Witten theory, the amplitudes with even higher number of gluons can in principle also be obtained.

In this work, we will extend certain aspects of \cite{Alday:2009yn} by considering the Pohlmeyer reduced theories describing spacelike minimal area worldsheets in AdS$_5$ and AdS$_4$.  We will explicitly construct the Lax pair for the reduced AdS$_5$ worldsheet, and recast the connection entirely in terms of the Lie algebra associated with the ${\rm A}_3={\rm D}_3$ root system (henceforth we will simply refer to this as the Lie algebra ${\rm A}_3={\rm D}_3$).  Restricting to AdS$_4$, the reduced system is equivalent to the affine Toda system for the Lie algebra ${\rm B}_2$, which can be thought of as a conformal gauge fixing of the conformal affine Toda system first considered in \cite{Babelon:1990bq, Aratyn:1990tr}.  For this case, we reformulate the flatness of the Lax connection of the  (conformal) affine Toda system as a system of Hitchin equations. We then show there exists a relation between the minimal area of the worldsheet and the Hitchin system moduli space K\"ahler potential associated to the affine Toda theory, generalizing the result found in \cite{Alday:2009yn}. This makes it possible in principle to evaluate scattering amplitudes embedded into AdS$_4$, given a construction of the related K\"ahler potential. The conformal extension of the affine Toda theory is essential for constructing solutions using the method pioneered by Leznov-Saveliev \cite{Leznov:1980tz}. We are able to recover the rotationally symmetric solutions of  \cite{Alday:2009yn} within this framework.

The paper is organized as follows.  In section 2, we apply Pohlmeyer reduction on the equations of motion for a spacelike worldsheet in AdS$_5$, and write it in a notation that is general for any AdS$_n$.  We then specialize to AdS$_5$ and write the Lax connection in terms of the Lie algebra ${\rm A}_3={\rm D}_3$.  In section 3, we recall how the AdS$_4$ reduced equations are related to the ${\rm B}_2$ affine Toda system. Using the Lax connection for the Toda equations, we show that it admits a reformulation in terms of Hitchin equations, whose moduli space K\"ahler potential we compute and find its derivative to be equal to the classical area of the worlsheet apart from a constant which we also find.  Section 4 is devoted to the construction of solutions, and we were able to reconstruct the solution to the radial Sinh-Gordon problem.
We then conclude with some remarks, and possible future directions.  In addition, we include 2 appendices: Appendix A we show how to construct the Lax pair for the (non reduced) worldsheet theories in AdS$_n$, and show that the dressing technique act transitively on the space of solutions; and in Appendix B we collect some details on explicit matrix representations of the involved Lie algebras.

\section{Reductions of equations}\label{reduction}
We first obtain the equations of motion for a string in AdS$_n$.  Recall that AdS$_n$ is given by the restriction
\be
-\left(Y^{-1}\right)^2-\left(Y^{0}\right)^2+\left(Y^{1}\right)^2\cdots +\left(Y^{n-1}\right)^2=-1
\ee
in ${\mathbb R}^{2,n-1}$.  We will avoid using the flat space ${\mathbb R}^{2,n-1}$ metric for computing the scalar product, and simply write ``$\cdot$''.  We implement the above restriction via a Lagrange multiplier $\Xi$ in the Polyakov action, giving
\be
S=\int dz d\zb \left(\pa Y \cdot \pab Y - \Xi\left(Y\cdot Y+1\right)\right).
\ee
Upon elimination of $\Xi$, the equations of motion read
\bea
\pa \pab Y^M &=& (\pa Y \cdot \pab Y) Y^M  \nn \\
Y\cdot Y &=& -1 \label{eomads}
\eea
and further, we have the Virasoro constraint
\be
\pa Y \cdot \pa Y = \pab Y \cdot \pab Y =0. \label{viraads}
\ee
This system of equations is integrable in the sense that it admits a Lax pair for arbitrary AdS$_n$ (see Appendix A).

For us, the most important piece of physical information is the worldsheet area.  Therefore, we need to keep track of the on-shell worldsheet Lagrangian density, hence we define
\be
\pa Y \cdot \pab Y = e^{2 \alpha(z,\zb)}.
\ee
We wish to construct a set of differential equations for $\alpha$.  To do so, we will need to introduce a complete set of vectors in AdS$_n$, which we do by defining a complete set of vectors in ${\mathbb R}^{2,n-1}$.  We take the vectors
\be
e=\begin{pmatrix} Y \\ \pa Y \\ \pab Y \\ B_0 \\ B_1 \\ B_2 \\ \cdot \end{pmatrix}
\ee
with $B_i\cdot Y=B_i \cdot \pa Y = B_i \cdot \pab Y =0$.  Here we have assumed that $\pa Y \cdot \pab Y \neq 0$ to ensure that $\pa Y$ is independent of $\pab Y$: this is equivalent to the statement that the worldsheet is non singular, and so $\alpha$ is finite.  We wish to consider spacelike surfaces in AdS, and so one of the (real) vectors $B_i$ must be timelike, and without loss of generality, we choose
\bea
B_i \cdot B_j = \eta_{ij} \label{bdotb} \\
\eta = \begin{pmatrix} -1 & 0 \\ 0 & {\mathbb I}_{(n-3)\times (n-3)} \end{pmatrix}.
\eea
These allow us to expand $\pa \pa Y (\pab \pab Y)$ in terms of the above basis and find
\bea
\pa \pa Y= (2\pa \alpha) \pa Y + u_i B_j \eta^{ij} \nn \\
\pab \pab Y= (2 \pab \alpha) \pab Y + \ub_i B_j \eta^{ij} \
\eea
with
\be
u_i \equiv B_i \cdot \pa \pa Y, \qquad \ub_i \equiv B_i \cdot \pab \pab Y.
\ee
The above expansion follows from the fact that $0=\pa(Y\cdot \pa Y)=Y\cdot \pa \pa Y$ and $0= \pa(\pa Y \cdot \pa Y)=2\pa Y \cdot \pa \pa Y$ (and so this requires the Virasoro constraint is met).
One may immediately prove that
\be
\pa \pa Y \cdot \pa \pa Y = u_i u_j \eta^{i j}= U(z)^2 \qquad \pab \pab Y \cdot \pab \pab Y = \ub_i \ub_j \eta^{i j}= \bar{U}(\zb)^2 \label{holo}
\ee
with $U(z)$ a holomorphic function, and $\bar{U}(\zb)$ an antiholomorphic function by directly differentiating both sides.  For the time being we leave both $U(z)$ and $\bar{U}(\zb)$ arbitrary.  Above, $U$ and $\bar{U}$ have been introduced as squares simply for later convenience.

We may also write
\bea
\pa B_i = -e^{-2 \alpha} u_i+\pa\beta_{i j}\eta^{j k} B_k  \\
\pab B_i = -e^{-2 \alpha} \ub_i+\pab \betab_{i j}\eta^{j k} B_k
\eea
with
\be
\pa \beta_{i j}\equiv \pa B_i \cdot B_j, \qquad  \pab \betab_{ij}\equiv \pab B_i \cdot B_j
\ee
(we take these as the definition of the symbol $(\pa \beta)_{ij}$ which we write in a loose notation $\pa \beta_{ij}$).  Note that $\pa \beta_{i j}=-\pa \beta_{j i}$ given relation (\ref{bdotb}).

We now restrict to AdS$_5$, although generalization to higher dimensional AdS is suggested by our calculation.  We use the above considerations to write
\bea
\pa e &=& \begin{pmatrix}
0 & 1 & 0 & 0 & 0 & 0 \\
0 & 2\pa \alpha & 0 & -u_0 & u_1 & u_2 \\
e^{2\alpha} & 0 & 0 & 0 & 0 & 0  \\
0 & 0 & -e^{-2\alpha} u_0 & 0 & \pa \beta_{01} & \pa \beta_{0 2} \\
0 & 0 & -e^{-2 \alpha} u_1 & \pa \beta_{01} & 0 & \pa \beta_{12} \\
0 & 0 & -e^{-2 \alpha} u_2 & \pa \beta_{02} & -\pa \beta_{12} & 0
\end{pmatrix}e \label{con1}\\
\pab e &=& \begin{pmatrix}
0 & 0 & 1 & 0 & 0 & 0  \\
e^{2\alpha} & 0 & 0 & 0 & 0 & 0  \\
0 & 0 & 2\pab \alpha & -\ub_0 & \ub_1 & \ub_2 \\
0 & -e^{-2\alpha} \ub_0 & 0 & 0 & \pab \betab_{01} & \pab \betab_{0 2} \\
0 & -e^{-2 \alpha}\ub_1 & 0 & \pab \betab_{01} & 0 & \pab \betab_{12} \\
0 & -e^{-2 \alpha} \ub_2 & 0 & \pab \betab_{02} & -\pab \betab_{12} & 0
\end{pmatrix}e. \label{con2}
\eea
The integrability condition for the two equations read
\bea
&&2 \pa \pab \alpha - e^{-2 \alpha}u_i \ub_j \eta^{i j} - e^{2 \alpha}=0 \label{alphaeq} \\
&&\pab (\pa \beta_{ij})- \pa (\pab \betab_{ij}) - \pa \beta_{i k} \pab \betab_{j m}\eta^{k m} + \pa \beta_{j k} \eta^{k m} \pab \betab_{i m}+e^{-2\alpha}\left(u_i \ub_j - u_j \ub_i\right)=0 \label{betaeq}
\eea
along with
\bea
\pab u_i = \pab \betab_{ij}\eta^{j k} u_k\equiv \pab \betab_{i}\,^j u_j \label{ueqn} \\
\pa \ub_i = \pa \beta_{ij}\eta^{jk } \ub_k \equiv \pa \beta_{i}\,^j \ub_j. \label{ubeqn}
\eea
The equation for $\alpha$ is only interesting when $U(z)\neq 0, \bar{U}(\zb)\neq 0$, and so we assume this henceforth.  Further, whatever the resulting equations from this point forward, it will always be possible to construct a Lax pair for the associated system.  The above is already stated as a ``zero curvature'' problem.  However, we need a one parameter family of connections to give the Lax pair.  This is easy to introduce because we note that equations are either homogeneous in $u_i$ or $\bar{u}_j$, or appear only in the combinations $u_i \bar{u}_j$.  This allows us to immediately see that the connection in (\ref{con1}) and (\ref{con2}) may be extended by replacing $u_i\rightarrow u_i \lambda$ and $\bar{u}_j\rightarrow \bar{u}_j/\lambda$.  This generalizes in a straightforward manner for all AdS$_n$ spaces.

We focus first on (\ref{ueqn}).  Given (\ref{holo}) we may write
\be
u_i = U(z) R_i \,^j v_j \rightarrow u= U(z) R v
\ee
with $v_j$ some set of real constants, and $R$ an $SO(1,2)$ matrix valued in ${\mathbb C}$: $R^T \eta R= \eta$ (in what follows, we will suppress the $SO(1,2)$ indices, and $\pa \beta_i\,^j\equiv \pa \beta$).  This allows us to write equation (\ref{ueqn}) as
\be
\left(R^{-1} \pab R - R^{-1} \pab \betab R \right) U(z) v=0
\ee
where we have multiplied in an $R^{-1}$, and in what follows, we divide by $U(z)$ as well.  Note first that $\pab \betab$ is valued in the Lie algebra of $SO(1,2)$, and therefore so is $R^{-1} \pab \betab R$.  The matrix $R^{-1}\pab R$ is also in the Lie algebra of $SO(1,2)$ and so full matrix multiplying $v$ is valued in the Lie algebra of $SO(1,2)$.  The only matrix that annihilates $v$ valued in the Lie algebra is
\bea
\vv &\equiv&  \begin{pmatrix} 0 & v_3 & -v_2 \\ v_3 & 0 & -v_1 \\ - v_2 & v_1 & 0 \end{pmatrix} \\
\vv v &=& 0.
\eea
Of course, any function $\pab \hat{\betab}$ multiplying $\vv$ would do as well, and so we find
\be
V v=v,\qquad V \equiv e^{\hat{\betab} \vv}.
\ee
For a general AdS, one would simply take the generators that leave $v$ invariant, and multiply them each by a separate $\pab \hat{\betab}$: these generators would generate the $SO$ group that leaves $v$ invariant.  The above allows us to conclude that
\be
\left(R^{-1} \pab R - R^{-1} \pab \betab R \right) = \pab \hat{\betab} \vv=V^{-1}\pab V= \pab V V^{-1}.
\ee
However, we note that $R$ is ambiguously defined.  In fact $R$ and $RV$ define the same $u$ via $u=U(z) R v=U(z) RV v$.  Hence, we replace $R$ by $RV$: this removes the right hand side of the above equation.  Therefore, we have shown that there exists an $R$ such that
\be
\pab \betab= \pab R R^{-1}, \qquad u=U(z) R v. \label{existR}
\ee
Similarly, one shows that there exists an $\bar{R}$ such that
\be
\pa \beta = \pa \Rb \Rb^{-1}, \qquad \ub= \bar{U}(\zb) \Rb v.
\ee
Because we have chosen the above $B_i$ to be real, one concludes that $\pa \beta^*=\pab \betab, U(z)^*= \bar{U}(\zb), R^*=\Rb$.  This makes the equations (\ref{alphaeq}) and (\ref{betaeq}) functions only of $\alpha$ and the complex matrix $R$, along with the auxiliary $U(z)$ and $\bar{U}(\zb)$ and $v_i$.  We may pick
\be
v=\begin{pmatrix} 1 \\ 0 \\ 0 \end{pmatrix}
\ee
without loss of generality.  Doing so reduces the equations to
\bea
2 \pa \pab \alpha &-& U(z) \bar{U}(\zb) e^{-2 \alpha}(\Rb v)^T \eta Rv - e^{2 \alpha}=0 \\
\pab (\pa \Rb \Rb^{-1})\! &- \!& \pa (\pab R R^{-1}) +[\pa \Rb \Rb^{-1}, \pab R R^{-1}]-U(z) \bar{U}(\zb)e^{-2\alpha}\left(Rv v^T \Rb^T \eta-\Rb v v^T R^T \eta \right)=0.  \nn \\
\eea
One may eliminate $U(z)$ and $\bar{U}(\zb)$ with a holomorphic change of coordinates and a change to the definition of $\alpha(z,\zb)$.  First, shift $\alpha(z,\zb)=\alpha'(z,\zb)+\frac14 \ln(U) + \frac14 \ln(\bar{U})$.  One may then multiply both of the above equations by $\frac{1}{U(z)^{\frac12}\bar{U}(\zb)^{\frac12}}$.  This eliminates $U(z)$ from all of the ``potential'' terms.  To remove $U(z)$ from the ``kinetic'' terms, one uses the change of variables $U(z)^{\frac12} dz=dw$.  We will use this to differentiate between the ``$z$-plane'' and ``$w$-plane'' in what follows: the functions $U$ and $\bar{U}$ only appear in equations when expressed in the $z$-plane, where as equations expressed in the $w$-plane have no occurrences of $U$ or $\bar{U}$.  We will switch to a somewhat bad notation and rename $\alpha'\rightarrow \alpha$ and $w\rightarrow z$.  Now, only the presence or lack of the functions $U,\bar{U}$ will denote whether we are in the $z$-plane or $w$-plane.

For completeness, we write the equations of motion in the $w$-plane
\bea
2 \pa \pab \alpha &-& e^{-2 \alpha}(\Rb v)^T \eta Rv - e^{2 \alpha}=0 \label{eqalpha2} \\
\pab (\pa \Rb \Rb^{-1})\! &- \!& \pa (\pab R R^{-1}) +[\pa \Rb \Rb^{-1}, \pab R R^{-1}]-e^{-2\alpha}\left(Rv v^T \Rb^T \eta-\Rb v v^T R^T \eta \right)=0. \label{eqbeta2} \nn \\
\eea
This equation generalizes to AdS$_n$ exactly as above, except that $R$ is valued in $SO(1,n-3;{\mathbb{C}})$, and $v$ is generalized appropriately, with the only non zero entry being a $1$ in the $0^{\rm th}$ position.  One must also generalize the argument preceding (\ref{existR}): the right hand side $\vv$ is now an arbitrary member of the Lie algebra $SO(n-3)$.  However, there always exists a $V$ in $SO(n-3)$ such that $V^{-1}\pab V=\vv$, namely $V=P\left[\exp(\int_0^{\zb} \vv(z,\zb') d\zb')\right]$ where the integration is done keeping $z$ constant, and $P$ is path ordering.  In this discussion, and the previous one, we have been a bit glib when introducing $V$ as a globally define object.  Again, we work locally and assume that $\int_C d\zb \pab\hat{\betab}(z,\zb)=0$ for any closed contour $C$ (again, treating $z$ as an independent constant parameter in the integration).  If this is violated around some point, then $V$ is defined on some appropriate Riemann surface: however, this will not affect the local field equations.

Next, note that when $R$ is real, (\ref{eqbeta2}) is solved and (\ref{eqalpha2}) collapses to the the usual sinh-Gordon equation.  When $R$ is real, $R$ simply rotates the real basis $B_i$ into another real basis $B'_i$.  This may be physically irrelevant: one can imagine taking a worldsheet that lives in AdS$_3$ and embed this in AdS$_n$, and then choose to rotate the basis $B_i$ differently at every point.  However, what is important for us is that this does not couple to the problem at hand.

In fact, one can immediately see that (\ref{eqbeta2}) is a field strength associated with the vector $A_z = \pa \Rb \Rb^{-1}$, $A_{\zb}=\pab R R^{-1}$ coupled to a potential term written in terms of $v$, $R$ and $\Rb$.  Further, $A$ transforms as a field strength under left multiplication by a real $SO(1,n-3;{\mathbb C})$ matrix; this also preserves the relation $R^*=\Rb$.  Further, the potential transforms the same way as the field strength because $R^{-1}=\eta R^T \eta$, and so we find that
\bea
&&\!\!\! \pab (\pa \Rb' \Rb'^{-1})-\pa (\pab R' R'^{-1}) +[\pa \Rb' \Rb'^{-1}, \pab R' R'^{-1}]-e^{-2\alpha}\left(R'v v^T \Rb'^T \eta-\Rb' v v^T R'^T \eta \right) \nn \\
&&= r\Bigg(\pab (\pa \Rb \Rb^{-1})-\pa (\pab R R^{-1}) +[\pa \Rb \Rb^{-1}, \pab R R^{-1}] \nn \\
&&\qquad \qquad \qquad \qquad \qquad -e^{-2\alpha}\left(Rv v^T \Rb^T \eta-\Rb v v^T R^T \eta \right)\Bigg)r^{-1} \nn \\
&& \qquad R'=r R, \qquad \Rb'=r \Rb, \qquad r^*=r, \qquad r^T \eta r=\eta.
\eea
It is also clear that (\ref{eqalpha2}) is invariant under such a transformation.  An arbitrary element of $SO(1,n-3;{\mathbb C})$ can be written as $e^{B_{ij} \tau^{i j}} e^{b_{ij} \tau^{i j}}$ where $B$ are real coefficients and $b$ are imaginary coefficients, so using the above transformation, one may eliminate the coefficients $B$.  Such a transformation eliminates half the degrees of freedom, leaving only the imaginary angles.  This may be expected because (\ref{eqbeta2}) is imaginary.  However, in what follows, we will find another decomposition more convenient.

We now specialize back to AdS$_5$, and so will consider $R$ in $SO(1,2;{\mathbb C})$.  In what follows, we will take the following as the generators $SO(1,2)$,
\bea
\tau^{01}= \begin{pmatrix} 0 & 1 & 0 \\ 1 & 0 & 0 \\ 0 & 0 & 0 \end{pmatrix},
\quad \tau^{02}= \begin{pmatrix} 0 & 0 & 1 \\ 0 & 0 & 0 \\ 1 & 0 & 0 \end{pmatrix},
\quad \tau^{12}= \begin{pmatrix} 0 & 0 & 0 \\ 0 & 0 & 1 \\ 0 & -1 & 0 \end{pmatrix},
\eea
which satisfy
\bea
\tau^{ij}=-\tau^{ji}, &&\quad  [\tau^{ij},\tau^{kl}]=-\left(\eta^{ik}\tau^{jl}+\eta^{jl}\tau^{ik}-\eta^{il}\tau^{jk}-\eta^{jk}\tau^{il}\right), \nn \\
&&\Tr\left({\tau^{ij} \tau^{kl}}\right)=-2\left(\eta^{ik}\eta^{kl}-\eta^{il}\eta^{jk}\right).
\eea
Further, we will need the relations
\bea
e^{\mp \beta_{01}\tau^{01}} \tau^{02} e^{\pm \beta_{01}\tau^{01}}=\cosh(\beta_{01}) \tau^{02} \mp \sinh(\beta_{01})\tau^{12} \nn \\
e^{\mp \beta_{01}\tau^{01}} \tau^{12} e^{\pm \beta_{01}\tau^{01}}=\cosh(\beta_{01}) \tau^{12} \mp \sinh(\beta_{01})\tau^{02} \nn \\
e^{\mp \beta_{02}\tau^{02}} \tau^{12} e^{\pm \beta_{02}\tau^{02}}=\cosh(\beta_{02}) \tau^{12} \pm \sinh(\beta_{02})\tau^{01} \nn \\
e^{\mp \beta_{12}\tau^{12}} \tau^{01} e^{\pm \beta_{12}\tau^{12}}=\cos(\beta_{12})\tau^{01} \pm \sin(\beta_{12})\tau^{02} \nn \\
e^{\mp \beta_{12}\tau^{12}} \tau^{02} e^{\pm \beta_{12}\tau^{12}}=\mp \sin(\beta_{12})\tau^{01} + \cos(\beta_{12})\tau^{01}. \label{conjugate}
\eea
The above relations allow us to write
\be
e^{-b_{0i}\tau^{0i}}=e^{-\hat{B}_{12}\tau^{12}}e^{-b\tau^{01}} e^{\hat{B}_{12}\tau^{12}}
\ee
where $b_{01}=b\cos(\hat{B}_{12}), b_{02}=b\sin(\hat{B}_{12})$: this simply states that $\tau^{0i}$ transforms as a vector under $\tau^{ij}$ transformations, and so we may pick a $\tau^{ij}$ transformation to align the $\tau^{0i}$ components to only be along $\tau^{01}$.  If we wish to have imaginary $b_{0j},b$, we simply take $b$ imaginary, and $B_{12}$ real (fitting with the earlier notation: $b$'s are imaginary, and $B$'s are real).

Given the above, we may write an arbitrary $SO(1,2;{\mathbb C})$ matrix
\bea
R &=&e^{B_{ij}\tau^{ij}}e^{-b_{0i}\tau^{0i}}e^{b_{12}\tau^{12}} \nn \\
&=& e^{B_{ij}\tau^{ij}}e^{-\hat{B}_{12}\tau^{12}}e^{-b\tau^{01}} e^{\hat{B}_{12}\tau^{12}}e^{b_{12}\tau^{12}} \nn \\
&=& e^{B_{ij}\tau^{ij}}e^{-\hat{B}_{12}\tau^{12}}e^{-b\tau^{01}} e^{\betab_{12}\tau^{12}}.
\eea
where $b$ is imaginary, $\beta_{12}$ is complex, and $B_{ij}$ are real.  Neither equation (\ref{eqalpha2}) nor (\ref{eqbeta2}) depend on the real matrix $e^{B_{ij}\tau^{ij}}e^{-\hat{B}_{12}\tau^{12}}$, and so we may drop this from our discussion.  Note that again we have eliminated half the degrees of freedom from $R$.  Hence we take
\be
R=e^{-b\tau^{01}} e^{\betab_{12}\tau^{12}}, \quad \Rb=e^{b\tau^{01}} e^{\beta_{12}\tau^{12}}
\ee
and plug these into equations (\ref{eqalpha2}) and (\ref{eqbeta2}) and use relations (\ref{conjugate})  to find
\bea
&& 2 \pa \pab \alpha + e^{-2 \alpha}\cosh(2b) - e^{2 \alpha}=0 \label{alphaeq2}\\
&& 2 \pa \pab b - \sinh(2b)\left(e^{-2\alpha}+\pa \beta_{12} \pab \betab_{12}\right)=0 \label{beq}\\
&& \pab\left(\pa \beta_{12} \cosh^2(b)\right)-\pa\left(\pab \betab_{12} \cosh^2(b)\right)=0 \label{beta1eq}\\
&& \pab\left(\pa \beta_{12} \sinh^2(b)\right)+\pa\left(\pab \betab_{12} \sinh^2(b)\right)=0. \label{beta2eq}
\eea

The last two equations are quite interesting.  Let us first define two vectors
\bea
&& \As1_z \equiv \As1\equiv\pa B_{12}=\pa \Re(\beta_{12}), \qquad \As1_{\zb}\equiv\Abs1\equiv \pab B_{12}=\pab \Re(\beta_{12}) \nn \\
&& \As2_z \equiv \As2\equiv\pa b_{12}=i\pa \Im(\beta_{12}), \qquad \As2_{\zb}\equiv\Abs2\equiv \pab b_{12}=i\pab \Im(\beta_{12}).
\eea
Note that $\As1$ is real while $\As2$ is imaginary when working in Cartesian coordinates.  Switching to Cartesian component form, equations (\ref{beta1eq}) and (\ref{beta2eq}) become
\bea
&& \pa_i \left[-i\epsilon^{ij}\left(\As1_{j}-i\epsilon_{jk}\As2^k\right)\cosh^2(b)\right]=0 \\
&& \pa_i \left[-i\epsilon^{ij}\left(\As2_{j}-i\epsilon_{jk}\As1^k\right)\sinh^2(b)\right]=0.
\eea
The above two equations state that we have two $U(1)$ vector fields both with zero field strength.  This means that they are pure gauge, and so we write
\bea
&&\left(\As1_{j}-i\epsilon_{jk}\As2^k\right)\cosh^2(b)=\pa_j \Lambda_1 \\
&&\left(\As2_{j}-i\epsilon_{jk}\As1^k\right)\sinh^2(b)=\pa_j \Lambda_2.
\eea
Note that from this
\bea
&&\tanh^2(b)\left(-i\epsilon^{ij}\right)\pa_j \Lambda_1=\delta^{ij}\pa_j\Lambda_2. \label{hodge}\\
\leftrightarrow && \tanh^2(b)\pa \Lambda_1=\pa \Lambda_2, \qquad \tanh^2(b)\pab \Lambda_1=-\pab \Lambda_2. \nn
\eea
This is crucially important: the above states that $\Lambda_1$ and $\Lambda_2$ are not independent.
There are several other useful relations, all following from the above.  First it is clear that $\Lambda_1$ is real, and $\Lambda_2$ is imaginary.  Further we find
\bea
&&\pa \Lambda_1=\left(\As1+\As2\right)\cosh^2(b) \\
&&\pab \Lambda_1=\left(\Abs1-\Abs2\right)\cosh^2(b) \\
&&\pa \Lambda_2=\left(\As1+\As2\right)\sinh^2(b) \\
&&\pab \Lambda_2=-\left(\Abs1-\Abs2 \right)\sinh^2(b)
\eea
and identify
\bea
&&\pa\left(\Lambda_1-\Lambda_2\right)=\pa \beta_{12}\rightarrow \Lambda_1-\Lambda_2=\beta_{12} \\ &&\pab\left(\Lambda_1+\Lambda_2\right)=\pab \betab_{12}\rightarrow\Lambda_1+\Lambda_2=\betab_{12}
\eea
and
\bea
&&\pa \left(\Lambda_1+\Lambda_2\right)=\pa \beta_{12}\cosh(2b) \\
&&\pab\left(\Lambda_1-\Lambda_2\right)=\pab \betab_{12}\cosh(2b).
\eea

We now rewrite equations (\ref{alphaeq2})-(\ref{beta2eq}) using only $\alpha, b, \Lambda_1$ given relation (\ref{hodge})
\bea
&& 2 \pa \pab \alpha + e^{-2 \alpha}\cosh(2b) - e^{2 \alpha}=0 \label{alphaeq3}\\
&& 2 \pa \pab b - \sinh(2b)\left(e^{-2\alpha}+\frac{\pa \Lambda_1 \pab \Lambda_1}{\cosh(b)^4}\right)=0 \label{beq3}\\
&& \pab\left(\pa \Lambda_1\right)-\pa\left(\pab \Lambda_1\right)=0 \label{beta1eq3}\\
&& \pab\left(\tanh^2(b)\pa \Lambda_1\right)+\pa\left(\tanh^2(b)\pab \Lambda_1\right)=0. \label{beta2eq3}
\eea
We have now reduced the problem to three equations for three degrees of freedom ($\alpha$ is real, $b$ is imaginary, and $\Lambda_1$ is real).  From now on we will drop the $_1$ index on $\Lambda_1$.  The three dynamical equations of motion are derivable from the following Lagrangian density (henceforth simply ``Lagrangian'')
\be
\mathcal{L}= \pa \alpha \pab \alpha+ \pa b \pab b+ \tanh^2(b) \pa \Lambda \pab \Lambda+\frac12\left(e^{-2\alpha}\cosh(2b)+e^{2\alpha}\right).
\ee
Note that this model is reminiscent of the models originally found in \cite{Pohlmeyer:1975nb} (by considering the dS$^3$ coset model without imposing the Virasoro constraint) and also those extensions found and studied in \cite{Bakas:1995bm}, with tangents multiplying kinetic terms in the Lagrangian.

Given the above identifications/definitions and the discussion after equation (\ref{ubeqn}), it becomes straightforward to construct the Lax pair for the above system.  We do so explicitly and find
\bea
\pa \Psi(z,\zb;\lambda)=A(z,\zb;\lambda) \Psi(z,\zb;\lambda) \\
\pab \Psi(z,\zb;\lambda)=\Ab(z,\zb;\lambda) \Psi(z,\zb;\lambda) \\
\eea
with
\bea
A(z,\zb;\lambda)=\begin{pmatrix}
0           &     \lambda             &      0      &      0                      &      0       &        0 \\
0           &  2\pa \alpha      &      0      &    -\lambda \cosh(b) &  -\lambda \sinh(b)  &        0 \\
\lambda e^{2\alpha} &  0 & 0 & 0 & 0 & 0 \\
0       &  0           &  -\lambda \cosh(b)e^{-2\alpha} & 0 &  \pa b  &    \frac{\sinh(b)}{\cosh(b)^2}\pa\Lambda \\
0       &  0           &  \lambda \sinh(b)e^{-2\alpha} & \pa b &  0  &    \frac{1}{\cosh(b)}\pa\Lambda \\
0       &  0           & 0 & \frac{\sinh(b)}{\cosh(b)^2}\pa\Lambda & -\frac{1}{\cosh(b)}\pa\Lambda  & 0
\end{pmatrix} \\
\Ab(z,\zb;\lambda)=\begin{pmatrix}
0           &     0             &      \frac{1}{\lambda}      &      0                      &      0       &        0 \\
\frac{1}{\lambda} e^{2\alpha} &  0 & 0 & 0 & 0 & 0 \\
0           &  0 & 2\pab \alpha       &    -\frac{1}{\lambda}\cosh(b) &  \frac{1}{\lambda} \sinh(b)  &        0 \\
0                 &  -\frac{1}{\lambda} \cosh(b)e^{-2\alpha}& 0 & 0 &  -\pab b  &- \frac{\sinh(b)}{\cosh(b)^2}\pab\Lambda \\
0                 &  -\frac{1}{\lambda} \sinh(b)e^{-2\alpha} &0 & -\pab b & 0  &    \frac{1}{\cosh(b)}\pab\Lambda \\
0                 & 0 & 0 & -\frac{\sinh(b)}{\cosh(b)^2}\pab\Lambda & -\frac{1}{\cosh(b)}\pab\Lambda  & 0
\end{pmatrix}.
\eea
In the above, we have used a constant gauge transformation to introduce the factors of $\lambda$ in front of the other potential terms $e^{2\alpha}$ and $1$ appropriately.  In such a form, only the terms leading to the ``potential'' in the problem get dressed with $\lambda$.

We may gauge transform the Lax connection so that we may write it in terms of the $6\times 6$ representation of the Lie algebra ${\rm A}_3={\rm D}_3$.  First, we rename
\be
b=\phi_1, \qquad \alpha=\phi_2.
\ee
The Cartan-Weyl basis for ${\rm D}_3$ is given by the Cartan generators $H_1, H_2, H_3$ and $E_{[i,j,k]}$ where the triplet $[i,j,k]$ has $i,j,k\in \{-1,0,1\}$ with $i^2+j^2+k^2=2$. We may write the connection in this language as
\bea
A &=& -\pa \phi_1 H_1-\pa\phi_2 H_2 \nn \\ &&+\frac{\lambda}{\sqrt{2}}\left(e^{\phi_1-\phi_2}E_{[1,-1,0]}+e^{\phi_2}E_{[0,1,-1]}+e^{\phi_2}E_{[0,1,1]}+e^{-\phi_1 -\phi_2}E_{[-1,-1,0]}\right)  \\
&&+\frac12\frac{\pa\Lambda}{\cosh(\phi_1)^2} \left(e^{-\phi_1}E_{[1,0,-1]}-e^{\phi_1}E_{[-1,0,1]}+e^{-\phi_1}E_{[1,0,1]}-e^{\phi_1}E_{[-1,0,-1]}\right)\nn \\
\Ab &=& \pab \phi_1 H_1+\pab\phi_2 H_2 \nn \\ &&+\frac{1}{\sqrt{2}\lambda}\left(e^{\phi_1-\phi_2}E_{[-1,1,0]}+e^{\phi_2}E_{[0,-1,1]}+e^{\phi_2}E_{[0,-1,-1]}+e^{-\phi_1 -\phi_2}E_{[1,1,0]}\right) \\
&&-\frac12\frac{\pab\Lambda}{\cosh(\phi_1)^2} \left(e^{-\phi_1}E_{[-1,0,1]}-e^{\phi_1}E_{[1,0,-1]}+e^{-\phi_1}E_{[-1,0,-1]}-e^{\phi_1}E_{[1,0,1]}\right). \nn
\eea
We may rewrite this in a compact form noting that the various exponentials and signs can be accounted for by the adjoint action of the Cartan subalgebra.  Further, we note that the set $\{[1,-1,0],[0,1,-1],[0,1,1]\}$ appearing in $A$ are a basis of simple roots, and $\{[-1,-1,0]\}$ is the negative of the highest level root for this basis (the negative of these appears in $\Ab$): this will lead to a nice connection to the ${\rm B}_2$ root system.  Next, we define the matrices
\bea
&&E_{+}=U(z)E_{[1,-1,0]}+E_{[0,1,-1]}+E_{[0,1,1]}+U(z)E_{[-1,-1,0]} \\
&&E_{-}=\bar{U}(\zb)E_{[-1,1,0]}+E_{[0,-1,1]}+E_{[0,-1,-1]}+\bar{U}(\zb)E_{[1,1,0]} \\
&&\hat{E}=E_{[1,0,-1]}-E_{[-1,0,1]}+E_{[1,0,1]}-E_{[-1,0,-1]} \\
&&\phi=\phi_1 H_1 + \phi_2 H_2
\eea
where we have reintroduced the holomorphic function $U(z)$ earlier removed by a shift of $\alpha$ and a coordinate transformation.  It is easy to check that $[E_{+},\hat{E}]=[E_{-},\hat{E}]=0$ given the commutation relations for the Lie algebra ${\rm A}_3={\rm D}_3$,  or using the explicit $6\times 6$ representation given in Appendix B.  Using these, the gauge transformed Lax connection may be written
\bea
A &=& -\pa \phi+\frac{\lambda}{\sqrt{2}}e^{{\rm ad}_{\phi}}E_{+} + \frac12\frac{\pa\Lambda}{\cosh(\phi_1)^2} e^{-{\rm ad}_{\phi}} \hat{E} \\
\Ab &=& \pab \phi+\frac{1}{\sqrt{2}\lambda}e^{-{\rm ad}_{\phi}}E_{-} + \frac12\frac{\pab\Lambda}{\cosh(\phi_1)^2} e^{{\rm ad}_{\phi}} \hat{E}.
\eea
We can easily see that when $\Lambda={\rm const}$ that we have simply taken the ${\rm B}_2$ affine Toda system and embedded it in the ${\rm A}_3={\rm D}_3$ affine Toda system in the natural way (one takes the ${\rm D}_3$ system, and sets $\phi_3=0$).  In the above form, the integrability condition simply refers back to the commutation relations for representations of the Lie algebra ${\rm A}_3={\rm D}_3$, and so one may use any representation one wishes (e.g. the $4\times 4$).

%
%
%
%

\section{The affine Toda system and Hitchin equations}\label{hitchin}

The equations above for the AdS$_4$ case reduces to the equations of motion of a (Euclidean) two dimensional affine Toda field theory, an integrable massive field theory associated to the extended root system of the Lie algebra ${\rm B}_2$. We shall only need the general properties of the (affine) Lie algebra in this section, and so relegate all detailed forms of the generators and fundamental representations of ${\rm B}_2$ to the Appendix B which interested readers should consult.

The generalized affine ${\rm B}_2$ Toda equations follow from the action for canonically normalized scalar fields $\phi_1,\phi_2$ with the generalized Toda potential
\be
V(\phi)={\frac12}(e^{2\phi_2}+{\frac12}U\bar Ue^{2\phi_1-2\phi_2}+{\frac12}U\bar Ue^{-2\phi_1-2\phi_2})
\ee
We follow \cite{Alday:2009yn}  and introduce the holomorphic function which determines the asymptotic behavior of the worldsheet  $p(z)=\partial^2Y\cdot\partial^2Y\equiv U(z)^2$. Like done there, we will also assume $p(z)$ is a polynomial of appropriate degree encoding the information about conformal invariant parameters, namely cross ratios, of the related scattering amplitude. Note the following important difference from the AdS$_3$ case: unlike in that case, $p(z)$ need not be the square of another holomorphic polynomial in general.

We recall it was a major virtue of the method proposed in \cite{Alday:2009yn} that one can find the area of the worldsheet by studying the K\"ahler potential of an auxiliary Hitchin system. This Hitchin system arises from splitting the flat Lax connection into the sum of a worldsheet gauge potential and a Higgs field term. While for non-affine, i.e. finite, Toda systems, their equivalence to Higgs bundles and hence Hitchin systems is partially understood in the context of W-gravity \cite{Bilal:1990wn, Gervais:1992gg, Aldrovandi:1993nz, Bonelli:2009zp}, no natural geometric relation is known for the affine Toda systems. \footnote{After the first version of this paper appeared, we were informed by the author of \cite{Kneipp:2008dc} that a similar connection between Hitchin system and affine Toda equations can be argued from a different point of view.} In this section, we will show that the flatness condition for the Lax pair of the {\textit{affine}} Toda system can be rewritten as a set of Hitchin equations. In fact, our formulation of the connection to Hitchin equations works for both the ordinary and the \textit{generalized} affine Toda equations, as well as the conformal affine Toda equations and their generalizations. The reason for this shall become clear shortly.

Using this relation, we will work out the K${\rm\ddot a}$hler potential on the (metriced) space of solutions to the ${\rm B}_2$ affine Toda equations and further show that a first derivative of the K${\rm\ddot a}$hler potential is identified with the area of the minimal surfaces up to an integer multiple of ${\frac{\pi}{2}}$, assuming the holomorphic function $p(z)$ introduced above is a polynomial. In deriving this Alday-Maldacena-like formula, the extra non-dynamical fields of the conformal affine Toda theory \cite{Babelon:1990bq, Aratyn:1990tr} play an essential role.

In the extended conformal affine Toda system, the Lax pair for the affine Toda equations can be written simply
\bea\label{catlax}
& &{\cal A}_z=\pa\varphi+e^{{\rm ad}\varphi}{\cal E}_+\nn\\
& &{\cal A}_{\bar z}=-\pab\varphi+e^{-{\rm ad}\varphi}{\cal E}_-
\eea
where the fields are now valued in the affine Lie algebra associated to ${\rm B}_2$, ie. the central extension of the loop algebra ${\rm B}_2\otimes\mathds{C}[t,t^{-1}]$ at level $\hat c$, whose more familiar name is the Kac-Moody algebra. To distinguish from the affine Toda theory, we have changed the notation for the dynamical fields from $\phi_i$ to $\varphi_i$. Apart from this notation change, the two subsets of fields are identified with each other. The Cartan part of the Lax connections are defined as
\bea
\varphi=\varphi_1H_1+\varphi_2H_2+\eta L_{0}+\xi\hat c
\eea
and
\bea
& &{\cal E}_+=U(z)E_{\a_1}^0+E_{\a_2}^0+U(z)E_{\a_0}^1\nn\\
& &{\cal E}_-=\bar U(\bar z)E_{-\a_1}^0+E_{-\a_2}^0+\bar U(\bar z)E_{\-a_0}^{-1}
\eea
where the subscripts label the corresponding simple root of the affine Lie algebra, and the superscripts gives the \textit{homogeneous} grading by $L_0=t{\frac{d}{dt}}$. Forgetting the factors of $U(z)$ and $\bar U(\bar z)$ we will get the Lax pair for the standard conformal affine Toda equations. When it's necessary to distinguish, we could put a $\hat{}$ on top of ${\cal E}_{\pm}$ for the connections containing the (anti-)holomorphic functions $U(z)$ and $\bar U(\bar z)$. Note that our ensuing discussions do not depend on which case we choose, in other words, the conclusions are valid both in the $z$-plane and $w$-plane. This happens for the trivial reason that we introduce $U(z)$ only in the holomorphic component of the Lax connection, and similarly for the anti-holomorphic counterpart.

Now we are ready to  show that the zero curvature equation for the above Lax pair is actually equivalent to Hitchin's self-duality equations on a Riemann surface. The key observation is that we can split the flat Lax connection ${\cal A}$ into a {\textit{gauge potential}} and a \textit{Higgs field}, as follows
\bea\label{split}
& &{\cal A}_z=A_z+\Phi_z,\quad A_z=\pa\varphi,\quad \Phi_z=e^{{\rm ad}\varphi}{\cal E}_+\nn\\
& &{\cal A}_{\bar z}=A_{\bar z}+\Phi_{\bar z},\quad A_{\zb}=-\pab\varphi,\quad \Phi_{\zb}=e^{-{\rm ad}\varphi}{\cal E}_-
\eea
where for example $e^{{\rm ad}\varphi}{\cal E}_+:=e^{\varphi}{\cal E}_+ e^{-\varphi}$. Under this decomposition, the flat connection equation translates into
\be
F_{z\bar z}+[\Phi_z,\Phi_{\bar z}]=0
\ee
which can be checked explicitly, noticing the following identities
\bea
& &\bar\pa(e^{{\rm ad}\varphi}{\cal E}_+)=[\bar\pa\varphi, e^{{\rm ad}\varphi}{\cal E}_+]\nn\\
& &\pa(e^{-{\rm ad}\varphi}{\cal E}_-)=-[\pa\varphi, e^{{\rm ad}\varphi}{\cal E}_-]
\eea
Based on these two same identities, one can see that the other two equations of the Hitchin system, namely
\be
D_z\Phi_{\bar z}=D_{\bar z}\Phi_z=0
\ee
are simply identities. This implies that the solution space to the (conformal) affine Toda equations naturally carries a hyperK${\rm\ddot a}$hler structure which it inherits as a subspace of the total solution space of the Hitchin equations. We now study this hyperK\"ahler structure using the metric due to Hitchin.

Following Hitchin, we can give a (infinite dimensional)  hyperK${\rm\ddot a}$hler metric to the space of solutions to the (conformally extended) affine Toda equations which as we saw above also solves Hitchin equations
(see also \cite{Alday:2009yn})
\be
\delta s^2=\int\,d^2z\Tr[-\delta A_z\delta A_{\bar z}+\delta \Phi_z\delta\Phi_{\bar z}]
\ee
where the minus sign was inserted for the manifestly non-unitary form of the gauge potential. To evaluate this metric on solutions of the full set of conformal affine Toda equations, for both the dynamical fields $\varphi_1,\varphi_2$ and the non-dynamical field $\xi$ \,\footnote{The field $\eta$ is known to be free, and decoupled, as can be seen by explicit calculations and hence may be set to vanish at any stage of the calculations. }, we
follow closely the steps of \cite{Alday:2009yn}. We will ignore subtleties of the global structure of this moduli space and leave such considerations for the future, since we find it reasonable to assume for each holomorphic polynomial $p(z)$ and its conjugate, a unique solution exists to the minimal surface equations (at least in AdS$_4$). In such a case, the degree of the polynomial is fixed by the number of cusps or equivalently the number of cross ratios \footnote{To be more precise, the number of cusps is $m+4$ where $m$ is the degree of the polynomial $p(z)$ which contains 2$(m-1)$ real free parameters, agreeing with the number of parameters uniquely determining positions of the cusps in AdS$_4$.}.

To eliminate gauge transformations of the solutions, we impose the orthogonality condition
\be
\delta s^2(\delta A_\m,\delta \Phi_\m; D_\m\epsilon)=0
\ee
for arbitrary transformation parameter $\epsilon$, which after partial integrations leads to the equation
\be\label{gaugetran}
-D_z\delta A_{\bar z}-D_{\bar z}\delta A_{z}+[\Phi_z,\delta \Phi_{\bar z}]+[\Phi_{\bar z},\delta \Phi_z]=0
\ee
However, under a generic deformation of the solutions, by $\delta A_\m$ and $\delta \Phi_\m$ which continues to satisfy the Toda equaitons, the above quantity can be evaluated, after some straightforward and somewhat lengthy calculations, to be
\bea
& & -D_z\delta A_{\bar z}-D_{\bar z}\delta A_{z}+[\Phi_z,\delta \Phi_{\bar z}]+[\Phi_{\bar z},\delta \Phi_z]\nn\\
&=& (\bar p\delta p-p\delta\bar p)\left\{-(e^{2\varphi_1-2\varphi_2}-e^{2\eta-2\varphi_1-2\varphi_2})H_1\right .\nn\\
& &\left . +(e^{2\varphi_1-2\varphi_2}+e^{2\eta-2\varphi_1-2\varphi_2})H_2-\hat c e^{2\eta-2\varphi_1-2\varphi_2}\right\}
\eea
This means that a compensating gauge transform is induced when one varies the solution infinitesimally. Notice due to the uniqueness of solutions, it is necessary to simultaneously vary also at least one of the holomorphic functions $U(z)$ or $\bar U(\bar z)$.  One can take the suitable gauge transformation to be
\be
A_\m\rightarrow A_\mu-\pa_\m\gamma,\quad \Phi_\m\rightarrow\Phi_\m+[\gamma,\Phi_\m]
\ee
where the gauge transformation parameter is  $\gamma=\gamma_1H_1+\gamma_2H_2+\gamma_3\hat c+\gamma_4L_0$. To compensate for (\ref{gaugetran}), the parameters must satisfy the following set of equations, taken to linear order in variations
\bea
& &2\pa\bar\pa\gamma_1-(e^{2\varphi_1-2\varphi_2}-e^{2\eta-2\varphi_1-2\varphi_2})(\bar U\delta U-U\delta\bar U)-U\bar U\left\{2(\gamma_1-\gamma_2)e^{2\varphi_1-2\varphi_2}\right .\nn\\
& &\left . +2(\gamma_1+\gamma_2-\gamma_4)e^{2\eta-2\varphi_1-2\varphi_2})\right\}=0\nn\\
& &2\pa\bar\pa\gamma_2+(e^{2\varphi_1-2\varphi_2}+e^{2\eta-2\varphi_1-2\varphi_2})(\bar U\delta U-U\delta\bar U)+U\bar U\left\{2(\gamma_1-\gamma_2)e^{2\varphi_1-2\varphi_2}\right .\nn\\
& &\left . -2(\gamma_1+\gamma_2-\gamma_4)e^{2\eta-2\varphi_1-2\varphi_2})\right\}-4\gamma_2e^{2\varphi_2}=0\nn\\
& &2\pa\bar\pa\gamma_3-e^{2\eta-2\varphi_1-2\varphi_2}(\bar U\delta U-U\delta\bar U)+2(\gamma_1+\gamma_2-\gamma_4)e^{2\eta-2\varphi_1-2\varphi_2}U\bar U=0\nn\\
& &2\pa\bar\pa\gamma_4=0
\eea
Comparing this set of equations to the linearized variation of the conformally extended affine Toda equations which we list below for completeness
\bea
& &2\pa\bar\pa\delta\varphi_1-(e^{2\varphi_1-2\varphi_2}-e^{2\eta-2\varphi_1-2\varphi_2})(\bar U\delta U+U\delta\bar U)-U\bar U\left\{2(\delta\varphi_1-\delta\varphi_2)e^{2\varphi_1-2\varphi_2}\right .\nn\\
& &\left . +2(\delta\varphi_1+\delta\varphi_2-\delta\eta)e^{2\eta-2\varphi_1-2\varphi_2})\right\}=0\nn\\
& &2\pa\bar\pa\delta\varphi_2+(e^{2\varphi_1-2\varphi_2}+e^{2\eta-2\varphi_1-2\varphi_2})(\bar U\delta U+U\delta\bar U)+U\bar U\left\{2(\delta\varphi_1-\delta\varphi_2)e^{2\varphi_1-2\varphi_2}\right .\nn\\
& &\left . -2(\delta\varphi_1+\delta\varphi_2-\delta\eta)e^{2\eta-2\varphi_1+2\varphi_2})\right\}-4\delta\varphi_2e^{2\varphi_2}=0\nn\\
& &2\pa\bar\pa\delta\xi-e^{2\eta-2\varphi_1-2\varphi_2}(\bar U\delta U+U\delta\bar U)+2(\delta\varphi_1+\delta\varphi_2-\delta\eta)e^{2\eta-2\varphi_1-2\varphi_2}U\bar U=0\nn\\
& &2\pa\bar\pa\delta\eta=0
\eea
we find the following identifications
\bea
& & \gamma_1=\delta\varphi_1-\bar\delta\varphi_1,\quad \gamma_2=\delta\varphi_2-\bar\delta\varphi_2,\quad \gamma_3=\delta\xi-\bar\delta\xi
\eea
whereas $\gamma_4 $ is free and will be set to zero. For the time being, we will still carry around the constant free field $\eta$, as it will be convenient for the following evaluation of the metric. The holomorphic and anti-holomorphic variations of dynamical fields above are understood to satisfy respectively the holomorphically and anti-holomorphically variations of the generalized conformal affine Toda equations. A useful observation is that
\be
\delta\varphi_1+\bar\delta\varphi_1,\quad \delta\varphi_2+\bar\delta\varphi_2,\quad \delta\xi+\bar\delta\xi
\ee
instead satisfy the full infinitesimal variation of the Toda equations under both holomorphic and anti-holomorphic variations, i.e. when $\delta U(z)$ and $\delta\bar U(\bar z)$ are both non-vanishing. They should be identified with the general infinitesimal deformations of the solutions, as we do henceforth.

With the above identifications, we can now finally evaluate the metric under infinitesimal deformations. Due to the orthogonal condition with respect to gauge transformations, we can simplify the calculation slightly and only evaluate instead
\be
\delta s^2=-\int\,d^2z\Tr[\delta A_z\delta A_{\bar z}-D_z\gamma D_{\bar z}\gamma]+\int\,d^2z\Tr[\delta \Phi_z\delta\Phi_{\bar z}-[\gamma,\Phi_z][\gamma,\Phi_{\bar z}]]
\ee
The result turns out to be quite simple and even maybe illuminating, given by
\bea\label{todamet}
\delta s^2&=&4\int\,d^2z\left\{e^{-2\varphi_2}\cosh2\varphi_1\delta U\delta\bar U+e^{-2\varphi_2}\sinh2\varphi_1(\bar U\delta U\bar\delta\phi_1+U\delta\bar U\delta\varphi_1)\right .\nn\\
& &\left . -e^{-2\varphi_2}\cosh2\varphi_1(\bar U\delta U\bar\delta\phi_2+U\delta\bar U\delta\varphi_2)\right\}
\eea
where we finally set the constant $\eta$ to be zero. It is gratifying that although the non-dyamical field $\xi$ determined by $\varphi_1,\varphi_2$ must be taken into account for consistency of the solutions, the resulting metric has no dependence on $\xi$. One notices that by setting $\phi_1={i\pi}$ we can immediately recover the result for Sinh-Gordon system obtained by Alday and Maldacena.
Following them, one naturally conjectures the K${\rm\ddot a}$hler potential to be
\be
{\frac14}K=\int\,d^2z\left\{\pa\varphi_1\bar\partial\varphi_1+\pa\varphi_2\bar\pa\varphi_2+{\frac12}(e^{2\varphi_2}+U\bar U e^{-2\varphi_2}\cosh2\varphi_1)\right\}
\ee
which is simply given by the reduction of the conformal affine Toda Lagrangian to the `dynamical'  fields.
It is easily checked, by taking ${\frac12}(\delta\bar\delta +\bar\delta\delta)K$ one reproduces the metric in (\ref{todamet}) as we should.

Now, specialize to the case of a holomorphic polynomial\footnote{To obtain real solutions, we set $\bar U(\bar z)$ and $U(z)$ as complex conjugates. } given by
\be
p(z)\equiv U(z)^2=\kappa^m\prod\limits_{i=1}^m (z-z_i)
\ee
where $\kappa$ is an overall scaling of the coefficients. One can see that the following derivative of the K${\rm\ddot a}$hler potential
\be
I:=\pa_{\kappa}K|_{\kappa=1}=2m\int\,d^2ze^{-2\varphi_2}\cosh2\varphi_1(p\bar p)^{1/2}
\ee
and so on shell
\bea\label{formula}
& &{\frac{1}{m}}I-A=\int\,d^2z\,2(e^{-2\varphi_2}\cosh2\varphi_1U\bar U-e^{2\varphi_2})=\int\,d^2z\pa\bar\pa\varphi_2=\int_{|z|\gg1}{\frac{1}{2i}}(dz\pa\varphi_2-d\bar z\bar\pa\varphi_2)\nn\\
& & ={\frac{\pi}{2}}m
\eea
which is fixed  by the number of cusps and independent on kinematics, and hence we find a generalization of the relation between minimal surface area and the moduli space K${\rm\ddot a}$hler potential found in the Sinh-Gordon case in \cite{Alday:2009yn}. As observed there, since the overall scaling of coefficients $\kappa$ can be absorbed homogeneously into the variables $z$ and $z_i$, an equivalent differential which can be used to define the integral $I$ is $\sum_i(z_i\pa_i+\bar z_i\bar\pa_i)$, which explicitly realizes the derivative as taken with respect to the cross ratios or the moduli of the defining worldsheet.
In summary, the kinematically richer scattering amplitudes which can be embed into AdS$_4$ subspaces of AdS$_5$ may now also be obtained, in principle, by studying the hyperK\"ahler potential for the related Hitchin systems using the relation (\ref{formula}) we found.

\section{`Solitons' and the radial solution}
In the previous section, we saw that the powerful observation of \cite{Alday:2009yn} can be extended at least to the minimal surface problem in AdS$_4$ . However, unlike in \cite{Alday:2009yn}, we currently do not have the luxury of a detailed understanding of the related Hitchin systems  at the level of \cite{Gaiotto:2009hg}. It may further be noted that even in that case, obtaining the hyperK\"ahler metric could involve solving some highly nontrivial integral equations. This is despite the successful global understanding of holomorphic symplectic structures and asymptotic form of the holomorphic Darboux coordinates achieved in the tour de force \cite{Gaiotto:2009hg}.  Hence it remains a meaningful route for solving the minimal surface problem to study the integrable systems of equations directly, which we note is again highly nontrivial.

We limit the scope of our work in this direction to the application of the formalism of \cite{Leznov:1980tz} and \cite{Mikhailov:1980my, Mansfield:1982sv}, combined with insights on the construction of soliton solutions based on vertex operator representation of the affine algebra studied first in for example \cite{Olive:1984mb, Olive:1985xw, Olive:1993cm}. We look for solutions in a particular scaling limit of the soliton class of solutions, with many details of our construction paralleling that in \cite{Bonora:2002ay} for the Sinh-Gordon case, and in fact the solutions will end up being identical to theirs, hence only giving a very special solution to the full AdS$_4$ system.



Since \cite{Olive:1992iu}, it was understood that soliton solutions to imaginary coupling affine Toda equations \cite{Hollowood:1992by} can be constructed using the Leznov-Saveliev method, by specifying exponentials of nilpotent elements of the affine Lie algebra, or Kac-Moody algebra.
To give some background of the formalism, we first review the Leznov-Saveliev method in a form suitable to conformal affine Toda equations. Our notations follow closely that of \cite{Babelon:1990bq}.  For general aspects of the vertex operator construction of Kac-Moody algebras, see \cite{Frenkel:1980rn, Segal:1981ap}.

As mentioned above, the soliton solutions are \textit{created} by a group element constructed from the affine algebra, in our case B$_2^{(1)}$. For convenience, we specify the group element in two different ways
\be
W=e^{-\varphi}g_1=e^{\varphi}g_2
\ee
where again $\varphi$ is the vector made from fields of the conformal affine Toda theory. Then in a highest weight state corresponding to an irreducible representation with weight vector $\lambda$ of the Kac-Moody algebra, we have
\be\label{sol}
\langle\lambda|g_2g_1^{-1}|\lambda\rangle=e^{-2\varphi\cdot\lambda}
\ee
Moreover, under a Gaussian decomposition of the group elements $g_1$, $g_2$ into the product of graded elements under the grading by $T_3+hL_0$
\be
g_1=e^{K_-}N_+M_-\quad,\quad g_2=e^{K_+}N_-M_+
\ee
where $h$ is the Coxeter number of the finite Lie algebra and $T_3=\rho^\vee\cdot H$ with $\rho^\vee$ the dual Weyl vector. Further, it can be shown that $M_+,K_+$ are holomorphic and $M_-,K_-$ anti-holomorphic. They are constrained by the first order differential equations
\bea\label{MM}
&& \pa M_+\cdot M_-^{-1}=-e^{-{\rm ad}K_+}{\cal E}_+\nn\\
&& \bar\pa M_-\cdot M_-^{-1}=-e^{-{\rm ad}K_-}{\cal E}_-
\eea
which guarantees that (\ref{sol}) gives solutions to the Toda equations.

The functions $K_{\pm}$ should be considered as `initial conditions' and explicitly expanding one has
\be
K_{\pm}=\varphi_1^{\pm}H_1+\varphi_2^{\pm}H_2+\xi^{\pm}\hat c+\eta^{\pm}L_0\, .
\ee
 In the basic and fundamental representations of the $\widehat {\rm B}_2$ Kac-Moody algebra at level one we have
\bea\label{gensol}
&& e^{-2\xi}=\langle0| M_+M_-^{-1}|0\rangle \,e^{\xi^+-\xi^-}\nn\\
&& e^{-2\varphi_1-2\xi}=\langle\Lambda_1| M_+M_-^{-1}|\Lambda_1\rangle\, e^{\varphi_1^+-\varphi_1^-+\xi^+-\xi^-}\nn\\
&& e^{-\varphi_1-\varphi_2-2\xi}=\langle\Lambda_2| M_+M_-^{-1}|\Lambda_2\rangle \,e^{{1\over2}\varphi_1^+-{1\over2}\varphi_1^-+{1\over2}\varphi_2^+-{1\over2}\varphi_2^-+\xi^+-\xi^-}
\eea
from which we find further
\be
e^{2\varphi_2}={\langle\Lambda_1| M_+M_-^{-1}|\Lambda_1\rangle\langle0| M_+M_-^{-1}|0\rangle\over \langle\Lambda_2| M_+M_-^{-1}|\Lambda_2\rangle^2}\, e^{-\varphi_2^++\varphi_2^-}
\ee

Taking `initial conditions' as in \cite{Bonora:2002ay}
\be
\varphi_2^+(z)=-\half \log p(z)\quad,\quad \varphi^-_2(\bar z)=\half \log\bar p(\bar z)\quad,\quad \eta^{\pm}=0
\ee
the equations for $M_{\pm}$ simplifies to
\bea
&& \pa M_+\cdot M_-^{-1}=-\sqrt{p(z)}\hat E_1\nn\\
&& \bar\pa M_-\cdot M_-^{-1}=-\sqrt{\bar p(\bar z)}\hat E_1^{\dagger}
\eea
where $\hat E_1=E_{\a_1}^0+E_{\a_2}^0+E_{\a_0}^1$ now contains no coordinate dependence. For this reason, the equations can be easily integrated, giving
\be
M_+=e^{-w\hat E_1}h_+ \quad,\quad M_-=e^{-\bar w\hat E_{-1}}h_-
\ee
with the coordinate $w$ defined by $dw=\sqrt{p(z)}dz$, and $h_{\pm}$ are constant group elements. The solutions are now given by
\be
\langle\lambda| M_+M_-^{-1}|\lambda\rangle=\langle\lambda|e^{-w\hat E_1}\,h_+ h_-^{-1}\,e^{\bar w\hat E_{-1}}|\lambda\rangle\, .
\ee
As can be seen, the constant group element $h_+ h_-^{-1}$ completely determines the solution.

To find the choices of $h_+ h_-^{-1}$ giving rise to solitons, it's easier to use an alternative basis for the ${\rm B}_2$ (and $\widehat {\rm B}_2$) algebra \cite{Fring:1991me}. Noticing the commutation relation between $\widehat E_1$ and its conjugate
\be
[\widehat E_1,\widehat E_{-1}]=\hat c
\ee
suggests that they are part of an infinite dimensional Heisenberg subalgebra (in other words oscillator algebra) of the Kac-Moody algebra, with the generators labeled by the exponents of $\widehat {\rm B}_2$, i.e. integers with values $4n+1$ and $4n+3$. We will not give the explicit formula for the higher indexed generators, as they differ from $\widehat E_{\pm 1}$ essentially only by the degree $4n$, and correspond to the class of roots $\a+n\delta$ where $\delta$ is the null root.
Under the Heisenberg subalgebra, the rest of the algebra (both ${\rm B}_2$ and $\widehat {\rm B}_2$) organize into orbits of the Coxeter element of the Weyl group, with the following commutation relations
\bea
& &[\widehat E_M,\widehat E_N]=\hat c M\delta_{M+N,0}\nn\\
& &[\widehat E_M, \widehat F^{\gamma_i}_N]=\gamma_i\cdot q([M])\widehat F^{\gamma_i}_{M+N}
\eea
where again the subscripts for $\widehat E_M$ belongs to the exponents of the algebra, which is not required for $\widehat F^{\gamma_i}_N$.
It can be recognized that $\widehat F^{\gamma_i}_N$ are the modes of operators which form eigenvectors of $\widehat E_M$.
The superscript of $\widehat F^{\gamma_i}_N$ is the representative element of an orbit under the Coxeter element, $\gamma_i=\epsilon(i)\a_i$ with the multiplicative factor $\epsilon(i)=\pm 1$ taken to be positive for the short root $\a_2$ and negative otherwise.  Explicitly, the Coxeter element is the product of the reflections by the two simple roots $\a_1,\a_2$ of ${\rm B}_2$
\be
\sigma=\sigma_2\cdot\sigma_1
\ee
resulting in two orbits with $h=4$ roots each, containing in order $(\a_2,\a_1+\a_2,-\a_2,-\a_1-\a_2)$ and $(-\a_1,\a_0,\a_1,-\a_0)$ respectively. Notice it's obvious from the roots in each orbit that
\be
\sum\limits_{p=1}^{h-1}\,\sigma^p=0\, .
\ee

The vector $q([M])$ is an eigenvector of the Coxeter reflection, with eigenvalue $\omega^{[M]}$ where $\omega=e^{2\pi i\over h}$ and $[M]=M | h$ is an exponent of ${\rm B}_2$, namely $1$ or $3$. Following \cite{Fring:1991me}, one finds the two corresponding eigenvectors are
\be
q(1)=\sqrt{2}e^{-i{3\pi\over8}}(\a_1+\a_2+i\a_2)\quad, \quad q(3)=\sqrt{2}e^{i{3\pi\over8}}(\a_1+\a_2-i\a_2)
\ee
and notice $q(1)=q(3)^*$. Also observe that the following orthonormal relations hold
\be
q(1)\cdot q(1)^*=h=4\quad,\quad q(1)\cdot q(3)^*=0\, .
\ee
In other words, we have a complete orthonormal basis
\be
q(\nu)\cdot q(\nu')^*=h\delta_{\nu,\nu'}\quad,\quad \sum_\nu q(\nu)q(\nu)^*=h\mathds{1}_{2\times2}\, .
\ee

We are now ready to construct the so called vertex operators for the elements of the Kac-Moody algebra,
\be
\widehat F^i(x)=\ell_i\sum\limits^\infty_{N=-\infty}\,x^{-N}\widehat F_N^{\gamma_i}
\ee
which physically corresponds to different `particle' species characterized by the eigenvalues $m_{i,M}=|\gamma_i\cdot q([M])|$
\be
[\widehat E_M, \widehat F^{i}(x)]=\gamma_i\cdot q([M])\,x^M\widehat F^{i}(z)
\ee
and the overall factor $\ell_i$ is determined by
\be
\langle0|\widehat F^i(x)|0\rangle={1\over4}\, .
\ee
It can be easily checked  that the construction
\be
\widehat F^i(x)={1\over4}\exp\left\{\sum\limits_{N>0}{\gamma_i^\vee\cdot q([-N])\,x^N\widehat E_{-N}\over N}\right\}\exp\left\{\sum\limits_{N>0}{-\gamma_i^\vee\cdot q([N])^*\,x^{-N}\widehat E_{N}\over N}\right\}
\ee
where the sum runs over all affine exponents of $\widehat {\rm B}_2$,  satisfies all the required commutation relations with $\widehat E_M$ for $\hat c=1$. Its (additive) mode expansion can be defined as usual using contour integrals in the complex plane, and further commutation relations of such modes will be omitted here.
Noticing the Coxeter element also rotates the basic and fundamental representations into each other, combined with its action on $\widehat F^i(x)$ this leads to the condition\footnote{The notation $\lambda_i^\vee$ stands for the co-weight vectors, defined as usual by ${2\over\a_i^2}\lambda_i$ for the fundamental representations.}
\be
\langle\lambda_j|\widehat F^i(x)|\lambda_j\rangle=e^{-2\pi\lambda_i^\vee.\lambda_j}\, .
\ee
which gives nontrivial phases to be included in the definition of $\widehat F^i(x)$ above, this fixes the overall normalizations of the operators $\widehat  F^i(x)$.
The Lie algebra ${\rm B}_2$ is non-simply laced and its vertex operator construction requires some care and the introduction of free fermion(s) \cite{Goddard:1986ts}, for further subtleties in the present context see also \cite{Kneipp:1993dn, Fring:1994mz, Kneipp:1999ch}.

To describe a mutli-soliton solution of the affine Toda equations involves normal ordering of the above vertex operators. From now on, we rename $V^i(x)=\widehat F^i(x)$ to emphasize its role as a  vertex operator, albeit a rather trivial one. The normal ordering of two such vertex operators is given by commutation relations  as follows
\be V^i(x_i)V^k(x_k)=X_{ik}(x_i,x_k)^{\hat c_j}:V^i(x_i)V^k(x_k):
\ee
where the overall phase factors in representation $|\lambda_j\rangle$ are included and explicitly
\bea\label{normal}
& &:V^i(x)V^k(y):={e^{-2\pi i(\lambda_i^\vee+\lambda_k^\vee)\cdot\lambda_j}}V^i_-V^k_-V_+^iV_+^k/4^2\nn\\
& & X_{ik}(x_i,x_k)=\prod\limits_{p=1}^4(1-\omega^{-p}{x_k\over x_i})^{\sigma^p(\gamma_i^\vee)\cdot\gamma_k^\vee}=\prod\limits_{p=1}^4(x_i-\omega^{-p}{x_k})^{\sigma^p(\gamma_i^\vee)\cdot\gamma_k^\vee}
\eea
with the postive/negative modes encoded in
\be
V_{\pm}^i={}\exp\left\{\sum\limits_{N>0}\mp{\gamma_i^\vee\cdot q([\pm N])\,x^{\mp N}\widehat E_{-N}\over N}\right\}
\ee
are the positive and negative modes. It can be seen from the second equation in (\ref{normal}) that the vertex operator $V^i(x)$ is nilpotent, its square vanishes. Notice an important difference from the simply-laced cases, here the normal ordering factor with regard to the exchange of $i$ and $k$ has the (anti-)symmetric property
\be
X_{i,k}(x_i,x_k)=(-1)^{|\a_i|^2\cdot|\a_k|^2}X_{k,i}(x_k,x_i)\, ,
\ee
manifesting the the fermionic behavior of the shorter root.

 A general n-soliton solution can be created by choosing the product of soliton creation operators $h_+h_-^{-1}=\prod\limits_{k=1}^{n}\exp{\{Q_kV^{i(k)}(x_k)\}}$, with the variables $\log Q$ and $\log x$ called respectively the coordinate and rapidity of the soliton. An interesting limit \cite{Bonora:2002ay} of the multi-soliton solutions is when one takes a certain kind of thermodynamic limit making the product of soliton operators infinite and continuously increasing in the rapidity. That such a limit exists may be understood from the related perspective of exact scattering matrices in integrable field theories \cite{Zamolodchikov:1978xm} which we will not further investigate here. We content ourselves with mentioning that in the thermodynamic limit, the solution of a set of nonlinear integral equations, called thermodynamic Bethe ansatz (TBA) \cite{Yang:1968rm, Zamolodchikov:1989cf} leads one precisely to the representation of solutions of affine Toda equations obtained below (see \cite{Fendley:1992dm, Fendley:1992jy, Cecotti:1992qh, Zamolodchikov:1994uw}).

 In ${\rm B}_2$ there are two kinds of vertex operators from the two simple roots with different lengths, as we already saw above they differ in the exchange symmetry.  We first consider the products of vertex operators all associated to $\a_1$, the longer root which we denote pictorially as $\circ$. A simple sample calculation is as follows
 \bea
& & \langle\lambda_j|e^{-w\widehat E_1}h_+h_-^{-1}e^{\bar w \widehat E_{-1}}|\lambda_j\rangle \nn\\
&=&  \langle\lambda_j|e^{-w\widehat E_1}\prod\limits_{k=1}^{n}\exp{\{Q_kV^{\circ}(x_k)\}}e^{\bar w\widehat E_{-1}}|\lambda_j\rangle\nn\\
&=&  \langle\lambda_j|e^{-w\widehat E_1}\prod\limits_{k=1}^{n}\{1+Q_kV^{\circ}(x_k)\}e^{\bar w\widehat E_{-1}}|\lambda_j\rangle\nn\\
&=& e^{-w\bar w}\langle\lambda_j|\prod\limits_{k=1}^{n}\{1+e^{2(wx_k+{\bar w\over x_k})}Q_kV^{\circ}(x_k)\}|\lambda_j\rangle
 \eea
 where we have used the freedom of rotation in the rapidity $x$-plane to absorb a factor of $-e^{i{3\pi\over8}}$ which keeps the normal ordering factors invariant. It will help to define the following notations
 \be
\Delta(x)=wx+{\bar w\over x},\quad x^-_{ij}=x_i-x_j,\quad x^+_{ij}=x_i+x_j\, .
 \ee
 Further expanding the above expression, one finds eventually
 \bea
 & & e^{-w\bar w}\langle\lambda_j|\prod\limits_{k=1}^{n}\{1+e^{2\Delta(x_k)}Q_kV^{\circ}(x_k)\}|\lambda_j\rangle\nn\\
 &=&e^{-w\bar w}\left\{1+\sum\limits_{k=1}^{n}\,{Q_k}e^{-2\pi i\lambda_{\circ}^\vee\cdot\lambda_j}e^{2\Delta(x_k)}+\sum\limits_{k<l}^{n}Q_kQ_le^{-4\pi i\lambda_{\circ}^\vee\cdot\lambda_j}e^{2\Delta(x_k)+2\Delta(x_l)}\left({x^-_{kl}\over x^+_{kl}}\right)^2\right .\nn\\
 &+&\left . \sum\limits_{k<l<m}^{n}Q_kQ_lQ_me^{-6\pi i\lambda_{\circ}^\vee\cdot\lambda_j}e^{2\Delta(x_k)+2\Delta(x_l)+2\Delta(x_m)}\left({x^-_{kl}\over x^+_{kl}}\right)^2\left({x^-_{km}\over x^+_{km}}\right)^2\left({x^-_{lm}\over x^+_{lm}}\right)^2+...\right\}\nn\\
 \eea
 where we rescaled $Q_k$'s by a factor of $4$. In the above manipulations, it is crucial to recall that $V^i(x_k)$'s are nilpotent and the exponentiated operator really terminates at the second order of expansion. Specializing to the basic and fundamental representations respectively, we have further
 \be\label{phases}
 e^{-2\pi i\lambda_{\circ}^\vee\cdot\lambda_0}=e^{-2\pi i\lambda_{\circ}^\vee\cdot\lambda_1}=1\quad, \quad e^{-2\pi i\lambda_{\circ}^\vee\cdot\lambda_2}=-1\, .
 \ee
 Shift $\Delta(x_k)=2(wx_k+{\bar w\over x_k}+{1\over2}\log Q_k)$, and take the thermodynamic limit, we can write the final results in terms of infinite dimensional Fredholm determinants
 \bea
& & \langle 0| M^\circ_+M_-^{\circ\,-1}|0\rangle=\langle\lambda_1| M^\circ_+M_-^{\circ\,-1}|\lambda_1\rangle=e^{-w\bar w}\det(1+W^\circ)\nn\\
& & \langle\lambda_2| M^\circ_+M_-^{\circ\,-1}|\lambda_2\rangle=e^{-w\bar w}\det(1-W^\circ)\nonumber\\
& & W^\circ_{kl}=2e^{\Delta(x_k)/2}{\sqrt{x_kx_l}\over x_k+x_l}e^{\Delta(x_l)/2}\, .
 \eea
 Plugging into (\ref{gensol}) this gives us what we shall call the long root solution
\bea
& &\varphi_2=\Tr\left(\log{1+W^\circ\over 1-W^\circ}\right)+{1\over4}\log(p\bar p)\nn\\
& & \varphi_1=-{1\over2}(\varphi_1^+-\varphi_1^-)
\eea
In the continuous scaling limit, the expressions found here through the quite lengthy construction is actually the well-known solution of the Painlev${\rm\acute{e}}$ III equation in disguise. As the continuous limit of this solution has been extensively discussed in the literature \cite{Jevicki:2009bv, Bonora:2002ay, Zamolodchikov:1994uw, McCoy:1976cd}, we will skip further details and merely state that by choosing proper numerical values for $Q_k$'s, we can find a solution satisfying the boundary conditions as required in \cite{Alday:2009yn}. It is helpful to note the rapidity variable $x$ used here corresponds to $e^{\theta}$ in those references.

While recovering the rotationally symmetric solution in AdS$_3$ is certainly encouraging, the method spelled out above does not seem to lead to new genuine AdS$_4$ solutions. First of all, considering the corresponding limit of products of the short root vertex operators leads to a trivial solution. This can be seen quite easily by noticing that the counterpart of phase factors in (\ref{phases}) will now all become equal to unity. Hence due to (\ref{gensol}), no details of the expansion of the infinite product of operators survive after taking the ratios and the solutions are simply constants which can be chosen properly to solve the Toda equations. Secondly, one may attempt considering an alternating product of mixed type of vertex operators and hope to find new solutions in a similar limit. Inspecting the structure of the expansion of such products gives no hint about how to construct a solution taking real values. It is possible what we are finding is an artifact of the special kind of initial conditions we have taken to make the integration of (\ref{MM}) easier. In view of this, it may well be helpful to first obtain some intuitions by numerical methods, such as the approximate solutions given by by series expansion in \cite{Jevicki:2009bv}.

\section{Future directions}

In conclusion, we outline several future lines of investigation that we find interesting.  One obvious possibility is to extend the result of section \ref{hitchin} to cover the full AdS$_5$ reduced system found in section \ref{reduction}, or even the arbitrary AdS$_n$ case.  Further, to actually extract the world sheet area, one would need the map from boundary data (the $k_i$ defining the null polygon) to moduli space data.  For the null Polygonal boundary conditions, formulating it as a stokes problem separating null ``wedge'' solutions may facilitate this.

It would be interesting to construct the soliton transformations for the AdS$_5$ case, and see if any new solutions could be obtained.  An important open problem is controlling the end result of soliton solution generating techniques: one may hope that this is possible given the success of such techniques in integrable gravitational setups \cite{Belinski:2001ph,Elvang:2007rd}.  Being able to generate solutions of interest is an interesting although nontrivial possibility.

Also, it is known that the full Pohlmeyer reduction for AdS$_5\times$S$^5$ is integrable: in this situation, an explicit form of the Lax pair, and group theoretic interpretation, would be useful.  It is clear that two copies (with appropriate analytic continuation for the S$^5$) of the result given here is not sufficient.  Throughout our computation, we have applied the Virasoro constraint to the AdS equations of motion directly, assuming that there is no profile of the string along the other compact directions (e.g. S$^5$).  Therefore, to get the full AdS$_5\times $S$^5$, one could repeat the calculations done here without imposing the Virasoro constraint.  This would yield the equations of motion for the AdS$_5$ factor, and then one could analytically continue to get a similar set of equations for the S$^5$.  The Virasoro constraint would then be a constraint relating the stress tensors of the two sectors of the full theory.

Finally, in view of the success of the TBA in obtaining the solution for the Sine-Gordon equations, it will be interesting to explore this possibility for the affine Toda equations we encountered in the AdS$_4$ case. In  the context of massive integrable deformations of two dimensional minimal models \cite{Fateev:1987zh}, the deformed theory is known to be described by the ADE series of (affine) Toda equations \cite{Eguchi:1989hs, Cecotti:1991me}. Following this philosophy, we may write down the S-matrices for the ${\rm B}_2$ affine Toda field theory as in \cite{Delius:1991kt} and try to find a thermodynamic quantity corresponding to the Painlev$\rm{\acute e}$ transcendent. It should be noted that the scattering matrices for non-simply laced affine Toda theories are `diagonal', while those used in the Painlev$\rm{\acute e}$ case were not. Although \cite{Zamolodchikov:1994uw} conjectures that worldsheet supersymmetry may not be essential, this distinction certainly deserves further understanding.

We look forward to addressing these problems in our future work.

\section*{Acknowledgements}
This work was supported under grants from NSERC of Canada and by the Perimeter Institute for Theoretical Physics.

\appendix

\section{Integrability of the AdS$_n$ worldsheet theory.}

In the text, we have already shown that the classical worldsheet equations of motion are
\bea
\pa \pab Y^M &=& (\pa Y \cdot \pab Y) Y^M  \nn \\
Y\cdot Y &=& -1 \label{eomadsA}
\eea
with the Virasoro constraint
\be
\pa Y \cdot \pa Y = \pab Y \cdot \pab Y =0. \label{viraadsA}
\ee
One may take the above (\ref{eomadsA}) and antisymmetrize in a $Y^N$ to find
\be
Y^{[N}\pa \pab Y^{M]} = 0. \label{eomcurrent}
\ee
This is in fact invertable, one may obtain (\ref{eomadsA}) from (\ref{eomcurrent}) by dotting in a $Y^N$, and using $Y\cdot Y=-1$.  Finally, we rewrite the above as
\be
\pa \left(Y^{[N} \pab Y^{M]}\right)+\pab \left(Y^{[N} \pa Y^{M]}\right) = 0 \label{eomcurrent2}
\ee
to write the above as a total divergence.

To relate this to a lax pair, we first define the matrix
\be
g\equiv Y\cdot \Gamma
\ee
where $\Gamma^M$ are the $SO(2,n-1)$ gamma matrices, satisfying
\be
\{ \Gamma^M, \Gamma^N\}=2\eta^{M N}.
\ee
Note also that
\be
g^2=-1, \qquad g^{-1}=-g, \qquad i\Gamma^{(-1)0}g^{\dagger}i\Gamma^{(-1)0}=-g^{-1}
\ee
One may then define the matrices
\bea
A\equiv -\pa g g^{-1} \qquad \bar{A}\equiv -\pab g g^{-1}.
\eea
Note that $\{\pa g, g\}=\{\pa g, g^{-1} \}=0$ due to $g^2=-1$.  Hence, $g^{-1} \pa g=-\Gamma_{NM}  Y^N \pa Y^M$, $g^{-1} \pab g=-\Gamma_{NM}  Y^N \pab Y^M$.  Finally, we see that the differential equation
\be
\pab A+\pa \bar{A}= -\left(\pa \left(Y_{[N} \pab Y_{M]}\right)+\pab \left(Y_{[N} \pa Y_{M]}\right)\right)\Gamma^{MN}=0 \label{laxeom1}
\ee
and there is a further identity
\be
\pab A - \pa \bar{A}-[A, \bar{A}]=0 \label{laxid}.
\ee
These two equations admit a Lax pair formulation
\be
\left(\pa + \frac{A}{1+\lambda}\right)\Psi=0, \qquad \left(\pab + \frac{\bar{A}}{1-\lambda}\right)\Psi=0. \label{laxpairA}
\ee
The integrability condition for the two equations is
\be
\left(\frac{(1-\lambda)\pab A-(1+\lambda)\pa \bar{A}-[A,\bar{A}]}{(1+\lambda)(1-\lambda)}\right)\Psi=0.
\ee
Further, we will restrict $i\Gamma^{(-1)0}(\Psi(z,\zb, \lambda^*))^{\dagger}i\Gamma^{(-1)0}=-\Psi(z,\zb, \lambda)^{-1}$, and so $\Psi$ is also invertible (this is done in analogy with the same restriction on $g$).  The integrability equation must be satisfied for all $\lambda$, and so both (\ref{laxeom1}) and (\ref{laxid}) must be satisfied by $A, \bar{A}$ because $A$ and $\bar{A}$ are independent of $\lambda$ and $\Psi$ is invertible.  Finally, we note that $\pa\Psi \Psi^{-1}=\frac{\pa g g^{-1}}{1+\lambda}$ (similarly for the $\pab$ equation) and so we may satisfy $\Psi|_{\lambda=0}=g$ as an initial condition on the solutions $\Psi$.

In the above $A, \bar{A}$ plays the role of a gauge connection, and so we see that (\ref{laxpairA}) is invariant under
\bea
\Psi&=&\chi \Psi' \nn \\
A &=& \chi A' \chi^{-1}- (1+\lambda)\pa \chi \chi^{-1} \nn \\
\Ab&=& \chi \Ab' \chi^{-1}- (1-\lambda)\pab \chi \chi^{-1}. \nn \\
\eea
We allow any such transformation $\chi(z,\zb,\lambda)$ such that $A'$ and $\Ab'$ are again independent of $\lambda$.  Such transformations are called the ``dressing'' solution generating technique \cite{Zakharov:1973pp} recently studied in \cite{Spradlin:2006wk}.

It is simple to see that this solution generating technique acts transitively on the space of solutions.  Suppose we have 2 solutions $g$ and $g'$.  We may construct the Lax pair, relevant $A$, $A'$, $\Ab$, $\Ab'$, and solutions $\Psi$, $\Psi'$.  Since $\Psi$ is invertible, $\Psi'\Psi^{-1}$ defines a good $\chi$: this $\chi$ brings us from the $\Psi$ solution to the $\Psi'$ solution, and is guaranteed to give an $A', \Ab'$ that are independent of $\lambda$ (plus whatever restrictions $\Psi'$ must satisfy).

\section{Representations for Lie algebras ${\rm A}_3={\rm D}_3$ and ${\rm B}_2={\rm C}_2$.}

We display here a possible representation of the Lie algebra ${\rm A}_3={\rm D}_3$.  For this we give first the Cartan generators
\bea
H_1=\begin{pmatrix}1 & 0 & 0 & 0 & 0 & 0 \\ 0 & -1 & 0 & 0 & 0 & 0 \\ 0 & 0 & 0 & 0 & 0 & 0\\
0 & 0 & 0 & 0 & 0 & 0\\0 & 0 & 0 & 0 & 0 & 0\\0 & 0 & 0 & 0 & 0 & 0\\ \end{pmatrix},&& \quad
H_2=\begin{pmatrix}0 & 0 & 0 & 0 & 0 & 0 \\ 0 & 0 & 0 & 0 & 0 & 0 \\ 0 & 0 & 1 & 0 & 0 & 0\\
0 & 0 & 0 & -1 & 0 & 0\\0 & 0 & 0 & 0 & 0 & 0\\0 & 0 & 0 & 0 & 0 & 0\\ \end{pmatrix}, \quad
 H_3=\begin{pmatrix}0 & 0 & 0 & 0 & 0 & 0 \\ 0 & 0 & 0 & 0 & 0 & 0 \\ 0 & 0 & 0 & 0 & 0 & 0\\
0 & 0 & 0 & 0 & 0 & 0\\0 & 0 & 0 & 0 & 1 & 0\\0 & 0 & 0 & 0 & 0 & -1\\ \end{pmatrix}
\eea
and then the generators associated with the positive simple roots
\bea
E_{[1,-1,0]}=\begin{pmatrix}0 & 0 & 1 & 0 & 0 & 0 \\ 0 & 0 & 0 & 0 & 0 & 0 \\ 0 & 0 & 0 & 0 & 0 & 0\\
0 & -1 & 0 & 0 & 0 & 0\\0 & 0 & 0 & 0 & 0 & 0\\0 & 0 & 0 & 0 & 0 & 0\\ \end{pmatrix},&& \quad
E_{[0,1,-1]}=\begin{pmatrix}0 & 0 & 0 & 0 & 0 & 0 \\ 0 & 0 & 0 & 0 & 0 & 0 \\ 0 & 0 & 0 & 0 & 1 & 0\\
0 & 0 & 0 & 0 & 0 & 0\\0 & 0 & 0 & 0 & 0 & 0\\0 & 0 & 0 & -1 & 0 & 0\\ \end{pmatrix} \nn \\
&& \kern-5em E_{[0,1,1]}=\begin{pmatrix}0 & 0 & 0 & 0 & 0 & 0 \\ 0 & 0 & 0 & 0 & 0 & 0 \\ 0 & 0 & 0 & 0 & 0 & 1\\
0 & 0 & 0 & 0 & 0 & 0\\0 & 0 & 0 & -1 & 0 & 0\\0 & 0 & 0 & 0 & 0 & 0\\ \end{pmatrix}
\eea
We further define
\bea
&& E_{[1,0,-1]}\equiv \left[E_{[1,-1,0]},E_{[0,1,-1]}\right],\nn \\ \quad
&&E_{[1,0,1]}\equiv -\left[E_{[1,-1,0]},E_{[0,1,1]}\right], \\
&&\quad E_{[1,1,0]}\equiv \left[E_{[1,0,-1]},E_{[0,1,1]}\right]=\left[\left[E_{[1,-1,0]},E_{[0,1,-1]}\right],E_{[0,1,1]}\right]. \nn
\eea
Of course any signs could be used above.  The above has defined all positive roots, and to define the negative roots we take $E_{[-i,-j,-k]}=\left(E_{[i,j,k]}\right)^T$, where $^T$ denotes the transpose.

The rank 2 Lie algebra ${\rm B}_2$  may be explicitly represented by matrices in the finite representations. The necessary information are encoded in the Cartan matrix
\be
\left(\begin{array}{cc} 2& -2\\ -1&2 \end{array} \right)
\ee
The isomorphic Lie algebra C$_2$ is obtained by transposing this Cartan matrix.

First, take the fundamental representation with weight vector $(1,0)$,  the Cartan subalgebra is generated by\footnote{We work with the Cartan-Weyl basis in this paragraph.}
\bea
H_1=\left(\begin{array}{ccccc} 1&0 &0 &0 &0\\ 0&0 &0&0&0 \\ 0&0 &0&0&0 \\ 0&0 &0&0&0 \\ 0&0 &0&0&-1\end{array} \right), \quad H_2=\left(\begin{array}{ccccc} 0&0 &0 &0 &0\\ 0&1 &0&0&0 \\ 0&0 &0&0&0 \\ 0&0 &0&-1&0 \\ 0&0 &0&0&0\end{array} \right)
\eea
The simple roots are $\alpha_1=e^1-e^2, \alpha_2=e^2$, and the affine root (which is minus the maximal root) $\alpha_0=-\alpha_1-2\alpha_2=-e^1-e^2$, where we have used $e^i$ to denote the unit vectors in Euclidean space. The corresponding matrices are
\bea
E_{\alpha_1}=\left(\begin{array}{ccccc} 0&1 &0 &0 &0\\ 0&0 &0&0&0 \\ 0&0 &0&0&0 \\ 0&0 &0&0&1 \\ 0&0 &0&0&0\end{array} \right), \quad E_{\alpha_2}=\left(\begin{array}{ccccc} 0&0 &0 &0 &0\\ 0&0 &\sqrt{2}&0&0 \\ 0&0 &0&\sqrt{2}&0 \\ 0&0 &0&0&0 \\ 0&0 &0&0&0\end{array} \right)
\eea
For negative roots, we use the convention $E_{-\alpha}=E_{\alpha}^{T}$. To get the right normalizations, we find
\be
E_{\alpha_0}=\left(\begin{array}{ccccc} 0&0 &0 &{0} &0\\ 0&0 &0&0&0 \\ 0&0 &0&0&0 \\ -1&0 &0&0&0 \\ 0&-1 &0&0&0\end{array} \right)
\ee

In the `non-unitary' gauge, the Lax connection for the ${\rm B}_2$ Toda system reads
\be
A_{z}=-2\partial\phi_i H_i+{\lambda/\sqrt{2}}\sum_{\alpha\in{\alpha_{1,2,0}}} E_{\alpha}\quad ,\quad A_{\bar z}={\lambda^{-1}/\sqrt{2}}\sum_{\alpha\in{\alpha_{1,2,0}}}e^{2\phi_{\alpha}}E_{-\alpha}
\ee
More explicitly, we have in the representation associated to the highest weight $(1,0)$
\bea
A_{z}&=&\left(\begin{array}{ccccc} -2\partial\phi_1&\lambda/\sqrt{2} &0 &0 &0\\ 0&-2\partial\phi_2 &\lambda&0&0 \\ 0&0 &0&\lambda&0 \\ -\lambda/\sqrt{2}&0 &0&2\partial\phi_2&\lambda/\sqrt{2} \\ 0&-\lambda/\sqrt{2} &0&0&2\partial\phi_1\end{array} \right), \nonumber\\
& &\nonumber\\
A_{\bar z}&=&\lambda^{-1}\left(\begin{array}{ccccc} 0&0 &0 &-e^{-2\phi_1-2\phi_2}/\sqrt{2} &0\\ e^{2\phi_1-2\phi_2}/\sqrt{2}&0 &0&0&-e^{-2\phi_1-2\phi_2}/\sqrt{2} \\ 0&e^{2\phi_2} &0&0&0 \\ 0&0 &e^{2\phi_2}&0&0 \\ 0&0 &0&e^{2\phi_1-2\phi_2}/\sqrt{2}&0\end{array} \right)
\eea
There is a common null vector of the Lax connections in this representation
\bea
&&\langle \Lambda_L| A_{\bar z}=(A_{z}+2\partial\phi_iH_i) | \Lambda_R\rangle=0,\nonumber\\
&& \langle \Lambda_L|\sim \left( \begin{array}{ccccc} e^{4\phi_1}\!&0&0&0&1\end{array}\right)\quad,\quad | \Lambda_R\rangle\sim \left( \begin{array}{c} 1\\0\\0\\0\\1\end{array}\right)
\eea
Notice that the null vector is gauge dependent.
Imitating the Leznov-Saveliev  construction one may find a solution for the field $\phi_1$, but not $\phi_2$ which is related to the worldsheet area.

There is a second fundamental representation associated with the fundamental weight $(0,1)$. The Cartan subalgebra is now given by
\bea
H_1=\left(\begin{array}{cccc} {\frac12}&0 &0 &0 \\ 0&{\frac12} &0&0\\ 0&0 &-{\frac12}&0 \\ 0&0 &0&-{\frac12} \end{array} \right), \quad H_2=\left(\begin{array}{cccc} {\frac12}&0 &0 &0 \\ 0&-{\frac12} &0&0\\ 0&0 &{\frac12}&0 \\ 0&0 &0&-{\frac12} \end{array} \right)
\eea
The admissible roots are given by
\bea
E_{\alpha_1}&=&\left(\begin{array}{cccc} 0 &0 &0 &0\\ 0&0 &1&0 \\ 0&0 &0&0 \\ 0&0 &0&0 \end{array} \right), \quad E_{\alpha_2}=\left(\begin{array}{cccc} 0&1  &0 &0\\ 0&0 &0&0 \\ 0&0 &0& 1 \\ 0&0 &0&0\end{array} \right),\nonumber\\
E_{\alpha_0}&=&\left(\begin{array}{cccc} 0 &0 &0 &0\\ 0&0 &0&0 \\ 0&0 &0&0 \\ 0&0 &0&1 \end{array} \right).
\eea
These lead to the Lax connections in explicit form
\bea
A_{z}&=&\left(\begin{array}{cccc} -\partial\phi_1-\partial\phi_2&\lambda/\sqrt{2} &0 &0 \\ 0&-\partial\phi_1+\partial\phi_2 &\lambda/\sqrt{2} &0 \\ 0&0 &\partial\phi_1-\partial\phi_2&\lambda/\sqrt{2} \\ \lambda/\sqrt{2} &0&0&\partial\phi_1+\partial\phi_2\end{array} \right), \nonumber\\
& &\nonumber\\
A_{\bar z}&=&\lambda^{-1}\left(\begin{array}{cccc} 0&0 &0 &e^{-2\phi_1-2\phi_2}/\sqrt{2} \\ e^{2\phi_2}/\sqrt{2}&0 &0&0 \\ 0&e^{2\phi_1-2\phi_2}/\sqrt{2} &0&0 \\ 0 &0&e^{2\phi_2}/\sqrt{2}&0\end{array} \right)
\eea
It is obvious in this explicit form that precisely due to the affine root, this Lax pair has no nontrivial common null vectors.

\end{document}